\documentclass[letter,twoside,twocolumn]{article}
\usepackage[super,sort&compress,comma]{natbib} 
\usepackage[version=3]{mhchem}
\usepackage{balance}
\usepackage{times,mathptm}
\usepackage{sectsty}
\usepackage{graphicx} 
\usepackage{lastpage}
\usepackage[format=plain,justification=raggedright,singlelinecheck=false,font=small,labelfont=bf,labelsep=space]{caption} 
\usepackage{fancyhdr}
\usepackage{graphicx}
\usepackage{setspace}

\usepackage{epstopdf}
\usepackage{subfigure}
\usepackage{amsmath}
\usepackage{amssymb}
\usepackage{amsthm}
\usepackage{amscd}
\usepackage[usenames]{color}

\def\th{\theta}

\begin{document}
\thispagestyle{plain}
\fancypagestyle{plain}{
%\fancyhead[L]{\includegraphics[height=8pt]{LH}}
%\fancyhead[C]{\hspace{-1cm}\includegraphics[height=20pt]{CH}}
%\fancyhead[R]{\includegraphics[height=10pt]{RH}\vspace{-0.2cm}}
\renewcommand{\headrulewidth}{1pt}}
\renewcommand{\thefootnote}{\fnsymbol{footnote}}
\renewcommand\footnoterule{\vspace*{1pt}% 
\hrule width 3.4in height 0.4pt \vspace*{5pt}} 
\setcounter{secnumdepth}{5}

\makeatletter 
\def\subsubsection{\@startsection{subsubsection}{3}{10pt}{-1.25ex plus -1ex minus -.1ex}{0ex plus 0ex}{\normalsize\bf}} 
\def\paragraph{\@startsection{paragraph}{4}{10pt}{-1.25ex plus -1ex minus -.1ex}{0ex plus 0ex}{\normalsize\textit}} 
\renewcommand\@biblabel[1]{#1}            
\renewcommand\@makefntext[1]% 
{\noindent\makebox[0pt][r]{\@thefnmark\,}#1}
\makeatother 
\renewcommand{\figurename}{\small{Fig.}~}
\sectionfont{\large}
\subsectionfont{\normalsize}

\fancyfoot{}
\fancyfoot[LO,RE]{\vspace{-7.5pt}}
\fancyfoot[CO]{\vspace{-7.2pt}\hspace{12.2cm}}
\fancyfoot[CE]{\vspace{-7.5pt}\hspace{-13.5cm}}
\fancyfoot[RO]{\footnotesize{\sffamily{1--\pageref{LastPage} ~\textbar  \hspace{2pt}\thepage}}}
\fancyfoot[LE]{\footnotesize{\sffamily{\thepage~\textbar\hspace{3.45cm} 1--\pageref{LastPage}}}}
\fancyhead{}
\renewcommand{\headrulewidth}{1pt} 
\renewcommand{\footrulewidth}{1pt}
\setlength{\arrayrulewidth}{1pt}
\setlength{\columnsep}{6.5mm}
\setlength\bibsep{1pt}

\vspace{1cm}

\twocolumn[
  \begin{@twocolumnfalse}
\noindent\LARGE{\textbf{Patterns on a Roll: A Method of Continuous Feed Nanoprinting}}
\vspace{0.6cm}

\noindent\large{\textbf{Elisabetta A. Matsumoto\textit{$^{a}$} and Randall D. Kamien,\textit{$^{b}$}}}\vspace{0.5cm}

%\noindent\textit{\small{\textbf{Received Xth XXXXXXXXXX 20XX, Accepted Xth XXXXXXXXX 20XX\newline
%First published on the web Xth XXXXXXXXXX 200X}}}

%\noindent \textbf{\small{DOI: 10.1039/b000000x}}
 \end{@twocolumnfalse} \vspace{0.6cm}
]

\noindent\textbf{Exploiting elastic instability in thin films has proven a robust method for creating complex patterns and structures across a wide range of lengthscales. Even the simplest of systems, an elastic membrane with a lattice of pores, under mechanical strain, generates complex patterns featuring long-range orientational order.  When we promote this system to a curved surface, in particular, a cylindrical membrane, a novel set of features, patterns and broken symmetries appears.  The newfound periodicity of the cylinder allows for a novel continuous method for nanoprinting.}
\section*{}
\vspace{-1cm}

\footnotetext{\textit{$^{a}$~Princeton Center for Theoretical Science, Princeton University, Jadwin Hall, Princeton, New Jersey, 08544, USA. Tel: 01-609-258-1143; E-mail: sabetta.matsumoto@gmail.com}}
\footnotetext{\textit{$^{b}$~Department of Physics and Astronomy, University of Pennsylvania, 209 S. 33rd Street, Philadelphia, Pennsylvania, 19103.}}

Again and again, periodicity enables the design and description of the materials we use to manipulate light, electrons, and other wave phenomena.  Elastic instabilities have long been utilized as a means of generating patterns in thin films
\cite{Bowden:1998p146,Chen:2004p597,Chan:2006p3238,Holmes:2007p3589,Katifori:2010p7635,Klein:2007p1116,Yin:2009p1470,Yin:2010p115402,Cao:2008p036102,Dressaire:2008p1198,Shin:2010p851}.  Generating, by self assembly, increasingly complex patterns on ever larger substrates with even finer structure remains a constantly receding goal.  Mastery of such processes would revolutionize the fabrication and design of novel materials with specific properties. 
Recently, several groups have exploited an elastic instability in a mechanically stressed elastic sheet perforated by a regular array of circular holes to generate intricate structures. 
Upon compression, the holes snap into elongated slits, revealing a diamond plate pattern featuring long-ranged orientational order \cite{Matsumoto:2009p021604,Zhang:2008p1192,Mullin:2007p084301},
 a robust mechanism acting from the macroscopic to the nanoscale.
This long-range order can be faithfully transmitted to substrates as printed patterns of metallic nanonparticles, resulting in optically interesting nanoscale materials.  
This method of one-step nano-assembly is limited by both the size of the elastic sheet and, more importantly, the limited number of possible symmetries that can be cast: there are five Bravais lattices in two-dimensions.  
To bypass these limitations, here we report on compression induced patterns arising from circular holes on a cylinder.  
The additional symmetry afforded us by the periodicity of the cylinder combined with the long-range elastic interactions between the slits leads to an extended library of motifs.
By capitalizing on the periodicity of the cylinder, one can upgrade from the ``sheet-at-a-time'' contact printing method for flat membranes to continuous-feed printing.  Thereby previously unachievable patterns can be transferred to arbitrarily long flat substrates with possibly interesting optical properties.

The collapse of holes in an elastic membrane -- both flat and curved -- occurs in the highly non-linear regime of elasticity, complicating the analysis.  Technological advances have made computational techniques, for instance finite element simulations, increasingly feasible. Only finite element simulations using specific models of nonlinear elasticity capture the entire process of the holes collapsing 
\cite{Mullin:2007p084301,Bertoldi:2008p2642}. 
These nonlinear effects are needed to predict the final shape of the collapsed holes and not merely their orientation. As the complexity of the system grows, such methods tend to be more accurate but lack the ability to distinguish between mechanisms responsible for different phenomena.  Analytic calculations generate intuitive solutions and can form the basis for subsequent numerical calculations.  With the sole assumption that each hole collapses to some elongated shape, our model uses only linear elasticity to successfully predict the orientational order in the diamond plate pattern and the herringbone pattern formed from an underlying triangular lattice.  Not only does our model shed light on the interactions in the system, it greatly facilitates the rational design of other patterns and devices.

\begin{figure*}[t!]
\includegraphics[width=17.1cm]{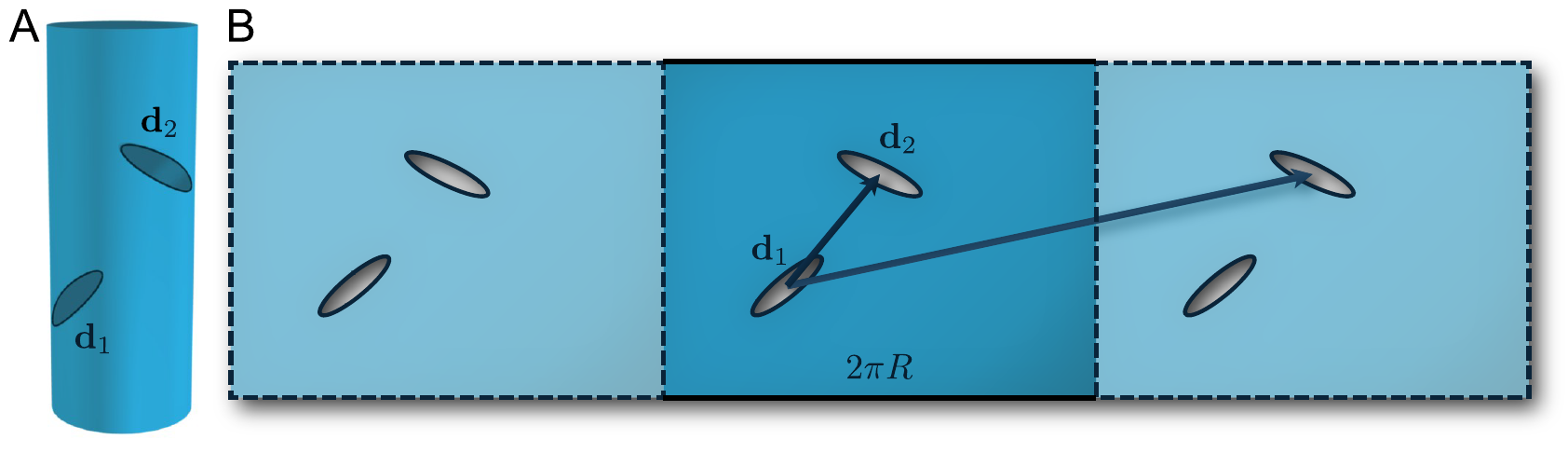}
\caption{({\bf A}) A cylinder of radius $R$ with two dislocation dipoles ${\bf d}_1$ and  ${\bf d}_2.$ ({\bf B}) The unrolled version of the cylinder with the first two image cylinders at $2 \pi n R \hat{\bf x},$ $n=\pm 1.$ }
\label{flattened cylinder}
\end{figure*}

Taking the theory of cracks as the starting point, the far field deformation of a collapsed hole can be approximated by a continuous distribution of parallel dislocations.  The details of the shape of the hole, and thus the exact distribution of dislocations, result from nonlinearities in the elasticity.  Since we consider only the final elongated slit, the first Fourier mode, a dislocation dipole of strength $b$ with dipole vector $\bf d$ described by the Burgers vector 
${\bf b}({\bf r})={\hat{\bf z}}\times {\hat{\bf d}} b \big(-\delta({\bf{r}}-{\bf d}/2)+\delta({\bf r}+{\bf d}/2)\big),$ 
sufficiently captures the deformation of a single hole.  Using linear elasticity theory, the interaction energy between a pair of dipoles with dipole vectors $\bf{d}_1$ and $\bf{d}_2$ centered at ${\bf r}=0$ and ${\bf r}={\bf R}$, respectively, is 
\begin{equation}\label{epair}
E=-\frac{Y_2 b^2 d_1 d_2}{\pi R^2} \left(\cos(\th_1+\th_2)\sin(\th_1)\sin(\th_2)+\frac{1}{4}\right),
\end{equation}
where $Y_2$ is the two dimensional Young's modulus and $\cos(\th_1)=\hat{\bf R}\cdot \hat{{\bf d}}_1$ and $\cos(\th_2)=\hat{\bf R}\cdot \hat{{\bf d}}_2$ are the angles each dipole makes with the vector connecting them.  For any arrangement of holes, we simply minimize the total energy -- the sum of all pairwise terms for a given set of holes -- for each of the dipole angles to find the resulting groundstate configuration upon hydrostatic compression.

The introduction of curvature immediately complicates the system.  Not only must the elastic energy for the membrane change to include curvature terms, the concept of a lattice is no longer well defined.  By adopting the topology of a cylinder, we may bypass these complications.  Because the cylinder is isometric to the plane, it has zero Gaussian curvature.  We may neglect additional bending energy if the thickness of the membrane is much less than all other relevant length scales.

In order to calculate the interaction between two holes on a mechanically compressed cylindrical membrane, we begin with the pairwise interaction energy, Eq. (\ref{epair}), of two holes on a flat membrane and consider the effect of the additional periodicity given by the cylinder.

Our geometry consists of an infinite elastic cylindrical shell of radius $R$ with its axis parallel to $\hat{\bf{z}}$ passing through the origin with two dislocation dipoles on it.  
Imagine cutting the cylinder along the line $r=R,\, \theta=0,$ and unrolling it onto the plane such that $x=R \theta, \, y=z.$  The two dipoles are now located at ${\bf d}_1=\{x_1,y_1\}$ and ${\bf d}_2=\{x_2,y_2\}.$ 
They interact not only through the shortest path and the first two replicas located at ${\bf d}_i\pm 2 \pi R \hat{\bf x}$, as they would if one simply considered periodic boundary conditions, but, due to the range of the potential, the interaction must also include terms from each of the infinite number of ``image" holes.  Note that the calculation on a cylinder is the same as the calculation on an infinite elastic sheet with a copy of the unrolled cylinder located every $2 \pi n R$ along the $\hat{{\bf x}}$ direction (see FIG. \ref{flattened cylinder}). Therefore, the reduced interaction energy, $\mathcal{E} = ER^2/(Y_2\pi b^2d_1d_2),$ between two dislocation dipoles on a cylinder is given by,
\begin{equation}\label{cyl energy sum}
\mathcal{E}=-\sum_{n=-\infty}^{\infty}\frac{\cos(\th_{1}+\th_{2}-2\th_n)\sin(\th_{1}-\th_n)\sin(\th_{1}-\th_n)+{\textstyle\frac{1}{4}}}{\big(X+2 \pi n\big)^2+Y^2},
\end{equation}
where $X=(x_2-x_1)/R$, $Y=(y_2-y_1)/R,$ and $\tan\theta_n=Y/(X-2 \pi n)$ is the angle between dipole 1 and the $n^{\rm th}$ image of dipole 2.  Upon expanding the trigonometric functions in equation (\ref{cyl energy sum}), the reduced energy consists of 5 terms, each of which may be summed,
\begin{equation}\label{sums}
\mathcal{E}=-\sum_{n=-\infty}^{\infty}\sum_{k=0}^{4}a_k(\theta_1,\theta_2)\frac{Y^{4-k}\big(X+2 \pi n\big)^k}{\big[(X+2\pi n)^2+Y^2\big]^3},
\end{equation}
where the $a_k(\th_1,\th_2)$ are given in Table \ref{aktable}.  We perform the tedious but straightforward sum in the appendix and derive the final expression for the energy.

\begin{figure}[h!]
\includegraphics[width=8.3cm]{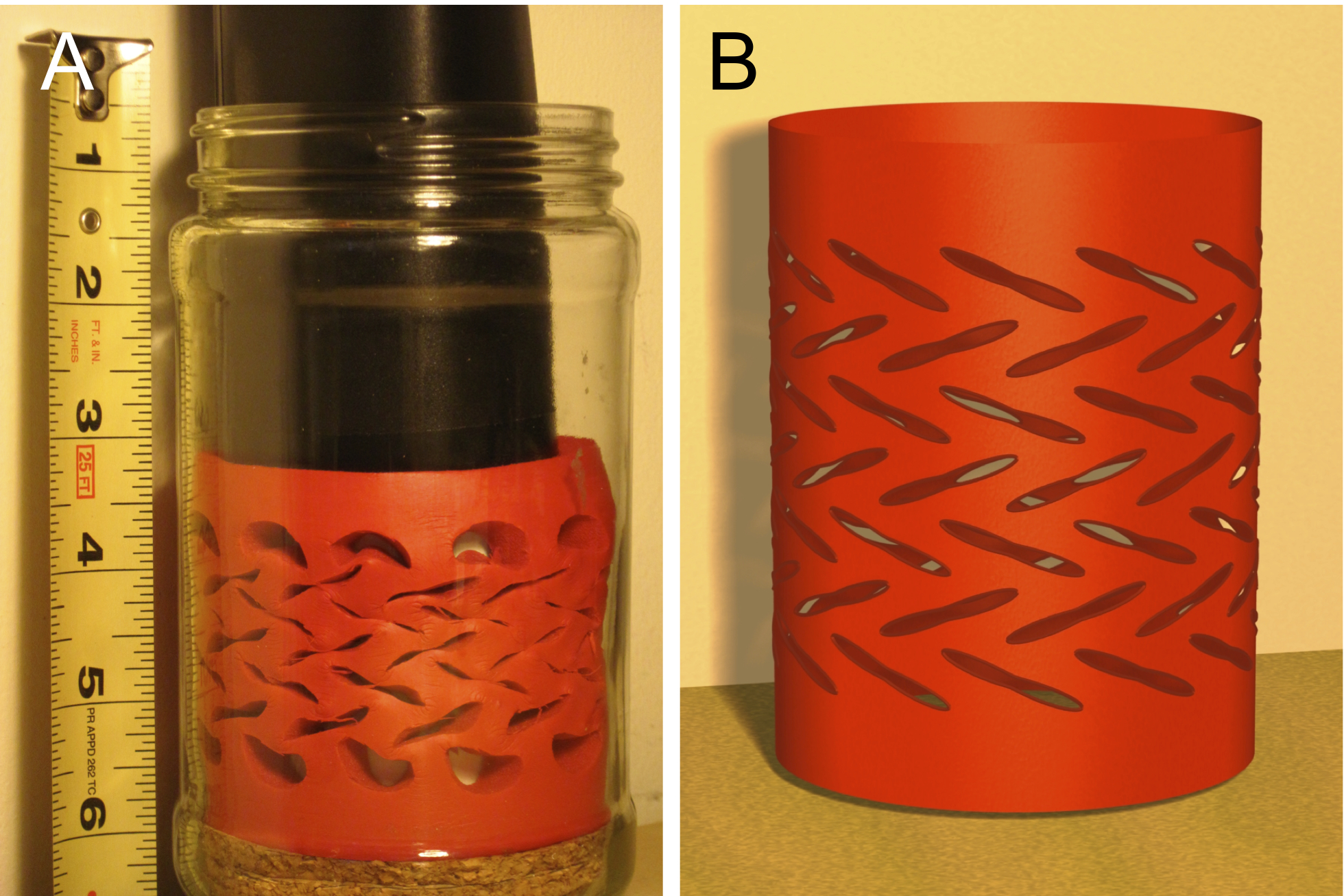}
\caption{The simple model using linear elasticity theory in equation (\ref{sums}) faithfully reproduces observed results for the square lattice (data on left).}
\label{square_fig}
\end{figure}

\begin{table}
\small
\caption{Expressions for the functions $a_k(\th_1,\th_2)$ used in the sum (\ref{sums}).}
\label{aktable}
\begin{tabular} {| c | l |}
\hline
$a_k(\th_1,\th_2)$ & \\
\hline
$k=0$ &$ -\cos(\th_1+\th_2)\cos\th_1\cos\th_2$\\
$k=1$ &$ -\frac{1}{2}\big[\sin2\th_1+\sin2\th_2+2\sin2(\theta_1+\theta_2)\big]$ \\
$k=2$ & $\frac{1}{2}\big[3\cos2(\theta_1+\theta_2)-1\big]$ \\
$k=3$ &$ -\frac{1}{2}\big[\sin2\th_1+\sin2\th_2-2\sin2(\theta_1+\theta_2)\big]$\\
$k=4$ &$ \cos(\th_1+\th_2)\sin\th_1\sin\th_2$ \\
\hline
\end{tabular}
\end{table}

The effect of long ranged elastic interactions serves to reinforce the same angular dipole interaction seen in the flat case for dipole pairs located at the same height on the cylinder, $E(X,Y=0)=-\frac{Y_2b^2d_1d_2\pi^2}{R^2}\cos(\th_1+\th_2)\sin\th_1\sin\th_2.$

Wrapping a lattice around a cylinder breaks the translational symmetry of the plane along one direction.  The orientation and magnitude of this periodic direction with respect to the lattice vectors and spacing, give new degrees of freedom with which to control patterns.  To gain a handle on this new phase space of possible patterns, we examine only those resulting from achiral lattice wrappings.

Before examining the plethora of complex lattices, it behooves us to consider the simple example of a square lattice, the results of which have been verified by a physical model (see Fig. \ref{square_fig}).  On a flat membrane, it is well known that a square lattice will produce a diamond plate pattern.  Although the pattern produced by the equivalent square lattice on a cylinder resembles a diamond plate pattern, the orientations of the collapsed holes are neither orthogonal to one another, nor do they lie along principal lattice directions.  

\begin{figure}[h!]
\includegraphics[width=8.3cm]{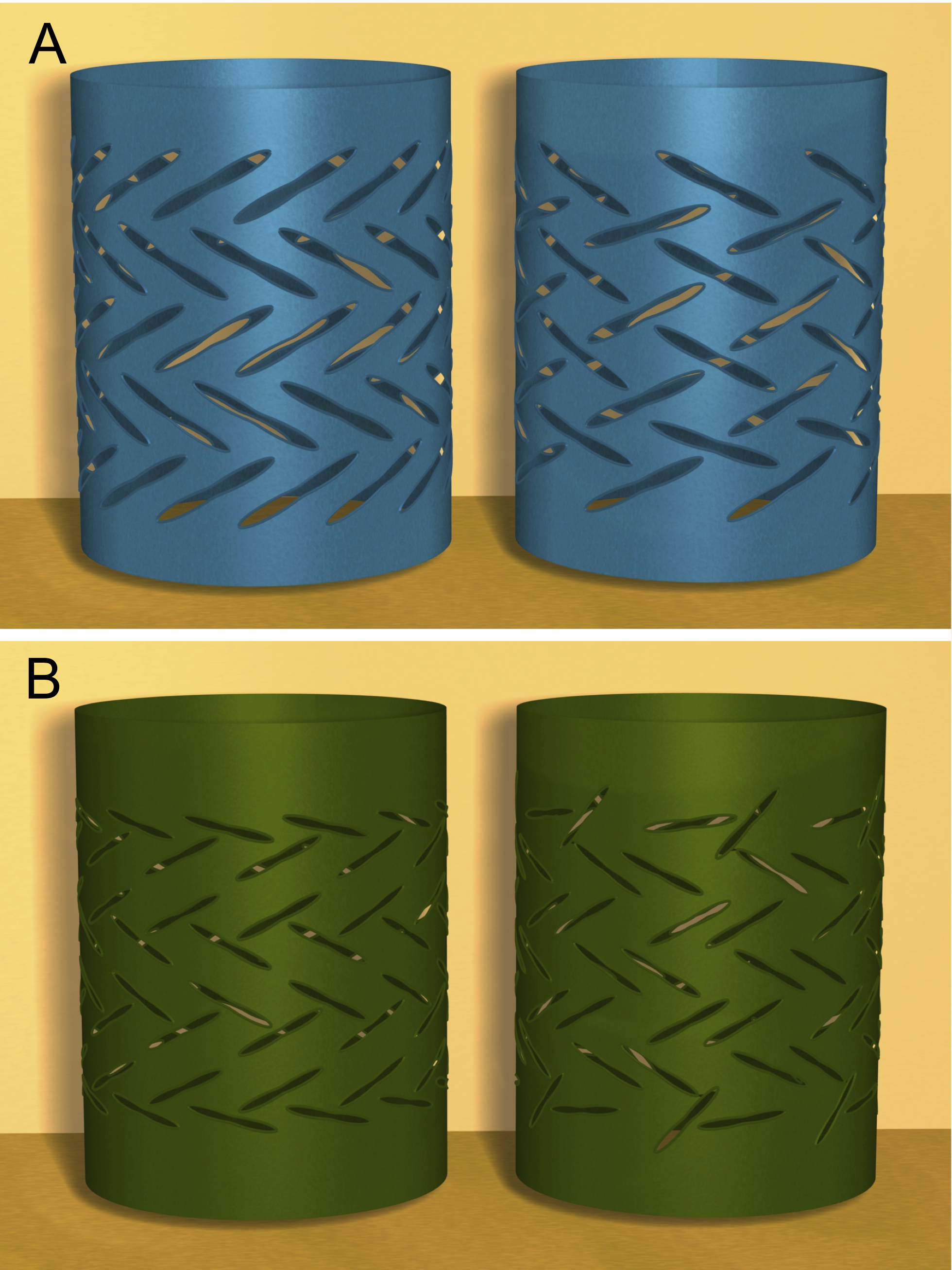}
\caption{The resulting patterns depend not only on the type of lattice but on its alignment with respect to the cylindrical axis, for example numerical results for triangular ({\bf A}) and honeycomb ({\bf B}) lattices.}
\label{tri_hc_fig}
\end{figure}

Fig.~\ref{tri_hc_fig} demonstrates the effect of changing the wrapping direction of simple lattices.  The triangular lattice (Fig. \ref{tri_hc_fig}A) produces stripes of alternating angles when one of its lattice vectors is aligned with the wrapping direction, and it produces a compressed diamond plate pattern when parallel to the cylindrical axis.  The honeycomb lattice, when wrapped in the two aforementioned directions (Fig. \ref{tri_hc_fig}B), produces very different results.  Of particular interest, the latter wrapping breaks chiral symmetry when the lattice is an odd number of hexagons tall.

A systematic study of variations of these simple lattices yields very precise control over the exact orientations of the dislocation dipoles for a few classes of patterns.  As variations on a theme, they can only add incrementally to our library of patterns. We must turn to increasingly complex lattices in order to broaden the range of accessible patterns.  

\begin{figure}[t!]
\includegraphics[width=8.3cm]{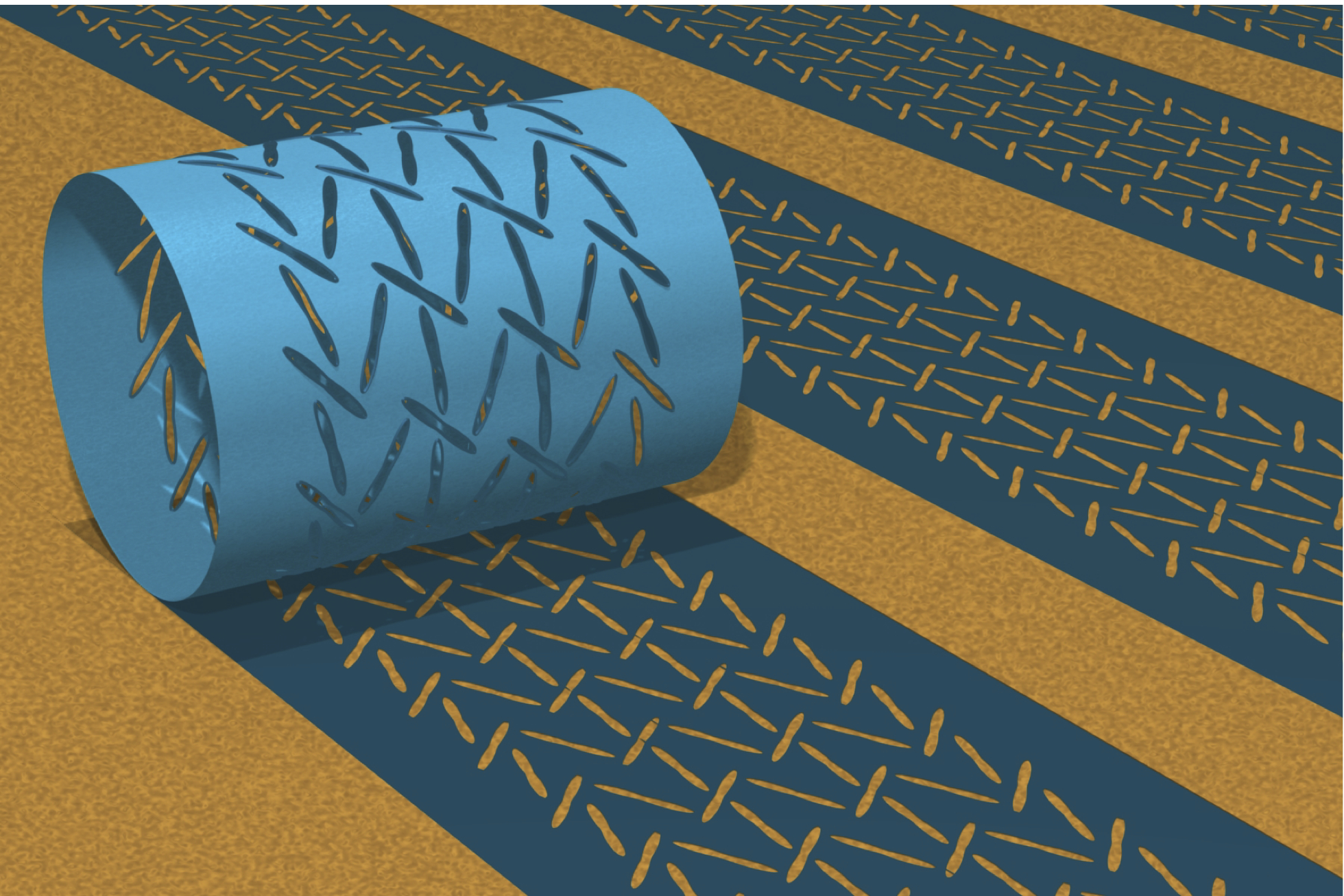}
\caption{An example application uses the pattern generated by the square lattice on a cylinder as a rolling printer onto a flat substrate.}
\label{printer_fig}
\end{figure}

The kagome lattice represents a perfect case study, as it has recently gained popularity in the condensed mater literature for its novel mechanical and vibrational properties \cite{Sun:2009p046811,Souslov:2009p205503,Mao:2011p011111}.  The kagome lattice consists of a triangular lattice ${\bf a}_1=\{2a,0\} \, {\bf a}_2=\{a,\sqrt{3}a\}$ with a basis of three points ${\bf b}_1=\{0,0\}, \, {\bf b}_2=\{a,0\}, \, {\bf b}_3=\{\frac{a}{2},\frac{\sqrt{3}a}{2}\}.$  The resulting pattern retains remnants of the symmetry of the underlying lattice, as shown in Fig. {\ref{printer_fig}}.

Varying the relative alignments of the lattice and cylindrical axis breaks the lattice symmetry and introduces a new degree of freedom to generate patterns. 
Moreover, unlike the diamond plate lattice on a flat membrane, the patterns on cylinders are sensitive to the number of holes both in circumference and in height.  
This immediately raises the question: how do we generate extensive patterns?  
Because cylinders are isometric to the plane, Gaussian curvature does not hinder the ability to transfer patterns onto a flat substrate.  
By capitalizing on the periodicity of the cylinder, one can upgrade from the ``sheet-at-a-time'' contact printing method outlined in \cite{Zhang:2008p1192} to continuous transfer. 
By filling the interior of the cylinder with an ink containing, for instance, nanoparticles, quantum dots, or polymers, the exact orientation and position of the holes may be transferred to a flat substrate of any dimension, shown in Fig. {\ref{printer_fig}}. The newly patterned surface adopts the same symmetries and optical properties of the cylindrical lattice not previously achievable with a flat membrane.   Likewise, by varying the ink pressure, changes in the pattern spacing and slit geometry can be controlled on the fly.
Once again, periodicity makes this possible.

\vspace{8pt}
\noindent{\large \bf Acknowledgements}\vspace{5pt}
\\
We acknowledge stimulating discussions with G.P. Alexander, T.C. Lubensky, and S. Yang.  This work was supported in part by NSF CMMI
09-00468 and the UPenn MRSEC Grant DMR11-20901.

\appendix
\section{Calculation of Geometric Sums}
In order to calculate the five sums in equation (\ref{sums}), we first note that {\small $\displaystyle s_{0}(p,q)= \sum_{n=-\infty}^\infty \frac{1}{(p+2 \pi n)^2+q^2}=\frac{\sinh q}{2 q(\cosh q-\cos p)}.$}  The five sums are given by:
%\begin{widetext}
\begin{eqnarray}
\sum_{n=-\infty}^\infty \frac{1}{\big[(X+2 \pi n)^2+Y^2)\big]^3}\hspace{-7pt}&=&\hspace{-7pt}\frac{1}{8 Y} \frac{\partial}{\partial Y}\bigg(\frac{1}{Y}\partial_y s_0(X,Y)\bigg)\nonumber\\
\sum_{n=-\infty}^\infty \frac{X+2 \pi n}{\big[(X+2 \pi n)^2+Y^2)\big]^3}\hspace{-7pt}&=&\hspace{-7pt}\frac{1}{8Y}\partial^2_{XY}s_0(X,Y)\nonumber\\
\sum_{n=-\infty}^\infty \frac{(X+2 \pi n)^2}{\big[(X+2 \pi n)^2+Y^2)\big]^3}\hspace{-7pt}&=&\hspace{-7pt}\frac{1}{8}{\partial_X^2}s_0(X,Y)-\frac{1}{8Y}{\partial_Y} s_0(X,Y)\nonumber\\
\sum_{n=-\infty}^\infty \frac{(X+2 \pi n)^3}{\big[(X+2 \pi n)^2+Y^2)\big]^3}\hspace{-7pt}&=&\hspace{-7pt}-\frac{Y}{8}{\partial^2_{XY}}s_0(X,Y)-\frac{1}{2}\partial_X s_0(X,Y)\nonumber\\
\sum_{n=-\infty}^\infty \frac{(X+2 \pi n)^4}{\big[(X+2 \pi n)^2+Y^2)\big]^3}\hspace{-7pt}&=&\hspace{-7pt}-\frac{Y^2}{8}{\partial_X^2}s_0(X,Y)\nonumber\\
&+&\hspace{-7pt} \frac{5 Y}{8}{\partial_Y}s_0(X,Y)+s_0(X,Y).
\end{eqnarray}

Combining these we arrive at a final, albeit complicated, expression for the  pairwise interaction of two dislocation dipole on a cylinder,
\begin{eqnarray}
E\hspace{-7pt}&=&\hspace{-7pt}-\frac{Y_2b^2d_1d_2}{4R^2}\frac{\pi}{H^3}\bigg[\frac{2H^2 \sinh Y}{Y}\nonumber\\
&+&\hspace{-7pt}2 H \sin X\sinh Y\big(\sin2\th_1+\sin2\th_2-\sin2(\th_1+\th_2)\big) \nonumber\\
&+&\hspace{-7pt}\big(3 H+\cos X\cosh2Y-\cos2X\cosh Y\big)\nonumber\\
&\times&\hspace{-7pt}\big(\cos2\th_1+\cos2\th_2-\cos2(\th_1+\th_2)\big)\nonumber\\
&+&\hspace{-7pt}Y\big((\cos 2X-3)\sinh Y+\cos X\sinh 2Y\big)\cos2(\th_1+\th_2)\nonumber\\
&+&\hspace{-7pt}Y\big((\cosh 2Y-3)\sin X+\cosh Y\sin 2X\big)\cos2(\th_1+\th_2)\bigg],\nonumber
\end{eqnarray}
where $H=\cos X-\cosh Y.$
%\end{widetext}

\footnotesize{
\bibliography{mybib2} %your .bib file
\bibliographystyle{rsc} %the RSC's .bst file
}

\end{document}